\documentclass[namedreferences]{solarphysics}

\usepackage[hyperref,optionalrh]{spr-sola-addons} 
\usepackage{graphicx}        
\usepackage{color}           
\usepackage{breakurl}        

\chardef\us=`\_

\begin{document}

\begin{article}
\begin{opening}

\title{Narrowband spikes observed during the 13 June 2012 flare in the 800-2000 MHz range}

\author[ addressref={aff1}]{\inits{M.}\fnm{Marian}~\lnm{Karlick\'{y}}\orcid{0000-0002-3963-8701}}
\author[ addressref={aff2}, corref, email={rybak@astro.sk}]{\inits{J.}\fnm{J\'{a}n}~\lnm{Ryb\'{a}k}\orcid{0000-0003-3128-8396}}
\author[ addressref={aff3}]{\inits{J.}\fnm{Jan}~\lnm{Ben\'a\v{c}ek}\orcid{0000-0002-4319-8083}}
\author[ addressref={aff1}]{\inits{J.}\fnm{Jana}~\lnm{Ka\v{s}parov\'a}\orcid{0000-0001-9559-4136}}

\address[id=aff1]{Astronomical Institute, Czech Academy of Sciences, 251 65 Ond\v{r}ejov,
Czech Republic}
\address[id=aff2]{Astronomical Institute, Slovak Academy of Sciences,
                  Tatransk\'{a} Lomnica, Slovakia}
\address[id=aff3]{Center for Astronomy and Astrophysics, Technical University of Berlin,
10623 Berlin, Germany}

\runningauthor{M. Karlick\'y, J. Ryb\'ak, J. Ben\'a\v{c}ek and J.
Ka\v{s}parov\'a} \runningtitle{Narrowband dm-spikes}

\begin{abstract}
Narrowband ($\sim$5 MHz) and short-lived ($\sim$0.01 s) spikes with three
different distributions in the 800-2000 MHz radio spectrum of the 13 June 2012
flare are detected and analyzed. We designate them as SB (spikes distributed in
broad band or bands), SZ (spikes distributed in zebra-like bands) and SBN
(spikes distributed in broad and narrow bands). Analyzing AIA/SDO images of the
active region NOAA 11504, a rough correspondence between groups of the spikes
observed on 1000\,MHz and peaks in the time profiles of AIA channels taken from
the flare sub-area close to the leading sunspot is found. Among types of spikes
the SZ type is the most interesting because it resembles to zebras. Therefore,
using auto-correlation and cross-correlation methods we compare SZ and SBN
spikes with the typical zebra observed  in the same frequency range. While the
ratio of SZ band frequencies with their frequency separation (220 MHz) is about
4, 5 and 6, in the zebra the frequency stripe separation is about 24 MHz and
the ratio is around 50. Moreover, the bandwidth of SZ bands, which consists of
clouds of narrowband spikes, is much broader than that of zebra stripes. This
comparison indicates that SZ spikes are generated different way than the zebra,
but similar way as SBN spikes. We successfully fit the SZ band frequencies by
the Bernstein modes. Based on this fitting we interpret SZ and SBN spikes as
those generated in the model of Bernstein modes. Thus, the magnetic field and
plasma density in the SZ spike source is estimated as about 79 G and 8.4
$\times$ 10$^{9}$ cm$^{-3}$, respectively.
\end{abstract}
\end{opening}

\section{Introduction}
     \label{S-Introduction}

Solar radio spikes are observed from tens of MHz up to GHz ranges
\citep{1977A&A....57..285D,1984SoPh...92..329K,1986SoPh..104..117S,
2016A&A...586A..29B,2014SoPh..289.1701M,2021ApJ...917L..32C}. They are
characterized by short duration (10 - 1000 ms) which decreases with increasing
frequency, narrow bandwidth ($\Delta$ f/f = 0.002 - 0.01) and the brightness
temperature up to 10$^{15}$ K \citep{1986SoPh..104...99B,2000A&A...354..287M}.
While \cite{2000A&A...354..287M} studied the spike parameters at two frequency
ranges (320-383 MHz and 873-1000 MHz), \cite{2014ApJ...789..152N} presented
results for the 1000-1500 MHz range that is the sub-range of the present study
(800-2000 MHz). Using the statistical methods they found the minimal and
maximal spike bandwidth as (according to their Figure 11): $(\Delta f)_{\rm
min} \sim$ 1 MHz and $(\Delta f)_{\rm max} \sim$ 100 MHz for 1000 MHz and
($\Delta f)_{\rm min} \sim $ 1 MHz and ($\Delta f)_{\rm max} \sim$ 10 MHz for
1500 MHz. The spikes are of high interest because their understanding can
provide detailed information about plasma processes in solar flares on kinetic
scales.

Among them, the decimetric-spikes belong to the most interesting because in
some cases they are recorded close to the starting frequency of type III bursts
and in relation to hard X-ray emissions \citep{2009A&A...504..565D}. As
concerns to their spatial localization, such observations are relatively rare.
For example, at 327 MHz and 410.5 MHz frequencies of the Nancay Radio
Heliograph \citep{1997LNP...483..192K}, clusters of spikes were observed above
the soft and hard X-ray flare sources \citep{2006A&A...457..319K}. Furthermore,
based on spikes observed in the 1 - 1.6 GHz range by Karl G. Jansky Very Large
Array (VLA), \cite{2021ApJ...911....4L} proposed that spikes are generated at
the termination shock formed above the flare arcade, where a diffuse
supra-arcade fan and multitudes of plasma downflows are present. This
interpretation is close to the idea that the narrowband dm-spikes are generated
by superthermal electrons in the magnetohydrodynamic turbulence in the magnetic
reconnection outflows \citep{1996SoPh..168..375K,2001A&A...379.1045B}.

On lower frequencies, the LOw Frequency ARray (LOFAR) recorded spikes in the 30
- 70 MHz range that were similar to individual Type IIIb striae observed in the
same event \citep{2021ApJ...917L..32C}. Authors estimated the spike emission
region of the order of $\sim$ 10$^8$ cm and brightness temperature as high as
10$^{13}$ K. For these spikes they suggested fundamental plasma emission
generated likely by weak/slow electron beams.

Several models of the spikes were suggested. In some of them, the runaway
electrons, accelerated in a direct-current electric field, were proposed
\citep{1981A&A...103..331K,1990ApJ...353..666T,1991ApJ...373..285W}. The
electron-cyclotron maser (ECM) mechanism, that directly generates
electromagnetic waves (spikes), was suggested by \cite{1982ApJ...259..844M},
\cite{1998PhyU...41.1157F}, \cite{2020ApJ...891L..25N} and
\cite{2017RvMPP...1....5M}. Moreover, \cite{1999ApJ...524..961S} and
\cite{2001A&A...379.1045B} presented the model, where spike frequencies
correspond to those of the upper-hybrid waves, and \cite{1996ApJ...467..465W}
the model with spike frequencies corresponding to the Bernstein modes.

There is an important additional aspect of the narrowband dm-spikes that is not
frequently considered in theoretical models. Namely, seeing them in radio
spectra, dm-spikes are in many cases clustered in bands (clouds) of spikes as
was for the first time shown by \cite{1994A&A...285.1038K}. The authors also
found that the frequencies of neighboring  bands are in the non-integer ratio
(1.06 - 1.54). In the paper by \cite{2021ApJ...910..108K} this result was not
only confirmed, but very narrow bands of spikes in the 7 November 2013 event
enable to successfully fit the band frequencies by the Bernstein modes.

In the model of the Bernstein modes the loss-cone distribution of
superthermal electrons together with much denser Maxwellian background plasma
is assumed. In this case, dispersion relations for the Bernstein modes can be
written as \citep{1975SoPh...43..431Z,2021ApJ...910..108K}
\begin{equation}
\epsilon_\parallel^{(0)} = 1 - 2\omega_\mathrm{pe}^2 \frac{e^{-\lambda}}{\lambda} \sum_{l=1}^{\infty} \frac{l^2 I_l(\lambda)}{\omega^2 - l^2 \omega_\mathrm{ce}^2 } = 0,
\label{eq1}
\end{equation}
\begin{equation}
\omega_\mathrm{pe}^2 = \frac{n_\mathrm{e} e^2}{m_\mathrm{e}
\epsilon_\mathrm{0}}, \qquad \lambda = \frac{k_\perp^2
v_\mathrm{tb}^2}{\omega_\mathrm{ce}^2},
\end{equation}
where $\epsilon_{\parallel}$ is the parallel component of the permittivity
tensor, $\epsilon_\mathrm{0}$ is the permittivity of free space,
$\omega_\mathrm{pe}$ is the electron plasma frequency, $n_\mathrm{e}$ is the
plasma density of the Maxwellian background plasma, $\omega_\mathrm{ce}$ is the
the electron cyclotron frequency, $\mathbf{k} = (k_\parallel, k_\perp)$ is the
wave vector parallel and perpendicular to the direction of the magnetic field,
respectively, $\omega$ is the wave frequency of the electrostatic wave,
$I_l(\lambda)$ is the modified Bessel function of $l$th order, $\lambda$ is the
dimensionless parameter, $m_\mathrm{e}$ is the electron mass, and $e$ is the
electron charge.

While the Maxwellian background plasma determines the dispersion relation of
the Bernstein modes, a generation of these electrostatic waves is owing to the
instability caused by electrons with the loss-cone distribution. The Bernstein
modes can be unstable if the double plasma resonance condition is fulfilled
\begin{equation}
\omega - \frac{k_\parallel u_\parallel}{\gamma_\mathrm{rel}} - \frac{s \omega_\mathrm{ce}}{\gamma_\mathrm{rel}} = 0,
\label{eq2}
\end{equation}
where $\gamma_\mathrm{rel} = (1 - v^2 / c^2)^{-\frac{1}{2}}$ is the
relativistic Lorentz factor and $s$ is the resonance gyro-harmonic number.
Their growth rate for some specific $(\omega,k_\perp)$ can be calculated as
\citep{1975SoPh...43..431Z}
\begin{equation}
\gamma(\omega, k_\perp) = - \frac{ \mathrm{Im} \, \epsilon_\parallel^{(1)}}
{\left[\frac{\partial \mathrm{Re} \, \epsilon_\parallel^{(0)}}
{\partial \omega}\right]_{\epsilon_\parallel^{(0)} = 0}}.
\label{eq3}
\end{equation}

We note that entirely the same equations are used for generation of the
upper-hybrid waves in the double-plasma resonance (DPR) model of solar radio
zebras \citep{2013SoPh..284..579Z}. In both these models it is assumed that the
Bernstein or upper-hybrid (electrostatic waves) modes are converted to the observed
electromagnetic waves by coalescence processes with the low-frequency waves or
mutual coalescence of the Bernstein modes. But, there is an essential
difference in these models. While in the model of the Bernstein modes the
Bernstein modes and escaping electromagnetic waves are produced in one source,
in the case of the zebra model with the upper-hybrid wave the zebra-stripes are
generated at different locations.

In this paper, after a brief statistical analysis of spikes observed in the 800-2000 MHz
range by the Ond\v{r}ejov radiospectrograph \citep{2008SoPh..253...95J}, we
present the narrowband spikes, observed during the 13 June 2012 flare. We study
these spikes because three different types of spike distributions were
detected in the radio spectrum of a single flare: a) spikes distributed in
broad band or bands, b) spikes distributed in zebra-like bands, and c) spikes
distributed in broad and narrow bands. We also search for their counterparts in
intensity variations in localized places of a flare in AIA/SDO images. We
compare the spikes in zebra-like bands with a typical zebra in the same
frequency range in order to distinguish their generation mechanisms. Finally,
we discuss the model of the presented spikes.

\section{Observations}

In Table~\ref{Table1} we present 18 spike events (consisting sometimes several
groups of spikes) observed in years 2007-2020 by the 800-2000 MHz Ond\v{r}ejov
radiospectrograph with the time and frequency resolution of 0.01 s and 4.7 MHz,
respectively \citep{2008SoPh..253...95J}. As can be seen here, these events are
always associated with flares (one flare classified as B flare, eight classified as C
flare, and nine classified as M flare). Most of the events were recorded before
the GOES (Geostationary Operational Environmental Satellites) soft X-ray flare
maximum (12 BM, see Table~\ref{Table1}), one before and during the maximum (1
BM, M), one before and after the maximum (1 BM, AM), one before and after the
maximum, but during bump on the GOES X-ray profile (1 BM, BAM), one after the
maximum, but during bump on the GOES X-ray profile (1 BAM), and two events
during the pre-flare phase (2 PF). As shown in Table~\ref{Table1}, most of
these spike events were associated with the hard X-ray (HXR) emission observed
by RHESSI or Fermi GBM instruments
\citep{2002SoPh..210....3L,2009ApJ...702..791M}. All these facts indicate that
spikes are generated during the impulsive phases of solar flares.

\begin{table}
  \caption{Spike events and corresponding hard X-ray (HXR) emission and GOES flares.
    HXR energy range corresponds to the highest range detected during the radio spike time interval.
    RHESSI data were the primary source for the HXR emission. If RHESSI data were not
    available, Fermi GBM data were used, see mark {\tiny f}.
    For the 04-Sep-14 event, RHESSI HXR data, denoted by {\tiny *},
    are available several seconds after the spike event, from 13:19 UT.
    Abbreviations in Notes
mean: BM spikes before the GOES maximum, M spikes at the maximum, AM spikes
after the maximum, BAM spikes during the bump on the GOES flare decay profile,
and PF spikes in the preflare phase.} \label{Table1}
\begin{tabular}{ccccccc}
  \hline
\multicolumn{2}{c}{Spike Event}  & \multicolumn{5}{c}{Flares (HXR, GOES)} \\
\hline
Time interval & Range & HXR & Start &  Max &  Class & Notes   \\
(UT) & (GHz) &  (keV) & (UT) &  (UT) &       &        \\
  \hline
12--Feb--10 09:40:40--09:41:20 & 0.8--1.1 & 25--50  & 09:38 & 09:42 &  B9.6 & BM \\
16--Feb--11 14:23:00--14:25:40 & 0.8--2.0 & 50--100   & 14:19 & 14:25 &  M1.6 & BM,M\\
18--Feb--11 10:13:30--10:14:30 & 0.8--2.0 & 50--100    & 09:55 & 10:11 &  M6.6 & AM \\
04--Mar--12 10:37:40--10:41:10 & 0.8--1.6 & 50--100${}^{\mbox{\tiny f}}$& 10:29 & 10:52 &  M2.0 & BM\\
13--Jun--12 13:02:40--13:30:00 & 0.8--1.8 & 50--100 & 11:29 & 13:17 &  M1.2 & BM,AM\\
14--Jun--12 10:39:30--10:42:30 & 0.8--2.0 & 12--25  & 10:45 & 10:50 &  C2.5 & PF\\
14--Jun--12 11:00:10--11:00:50 & 0.8--2.0 & no data  & 11:05 & 11:12 &  C5.0 & PF\\
11--Apr--13 06:56:00--07:02:00 & 0.8--2.0 & 50--100  & 06:55 & 07:16 &  M6.5 & BM\\
07--Nov--13 12:26:00--12:28:30 & 0.8--2.0 & no data  & 12:22 & 12:29 &  C5.9 & BM\\
08--May--14 10:02:10--10:04:00 & 0.8--2.0 & 100--300${}^{\mbox{\tiny f}}$ & 09:59 & 10:07 &  M5.2 & BM\\
09--Jun--14 12:27:10--12:37:00 & 0.8--2.0 & 50--100  & 12:24 & 12:29 &  C9.0 & BM,BAM\\
11--Jun--14 05:41:10--05:42:00 & 0.8--1.6 & 12--25  & 05:30 & 05:34 &  M1.8 & BAM\\
11--Jun--14 07:11:40--07:12:00 & 0.8--2.0 & 25--50  & 07:09 & 07:12 &  C2.8 & BM\\
12--Jun--14 09:33:50--09:36:50 & 0.8--1.3 & 100--300${}^{\mbox{\tiny f}}$ & 09:23 & 09:37 &  M1.8 & BM\\
12--Jun--14 15:59:30--15:59:50 & 0.8--1.6 & 25--50  & 15:57 & 16:03 &  C7.8 & BM\\
04--Sep--14 13:17:50--13:18:50 & 1.1--1.8 & 12--25${}^{\mbox{\tiny *}}$ & 13:10 & 13:30 &  C6.3 & BM\\
12--Sep--14 09:31:20--09:32:00 & 1.0--1.4 & 14--25${}^{\mbox{\tiny f}}$ & 08:24 & 09:55 & C3.2 & BM\\
22--Aug--15 06:41:40--06:42:50 & 0.8--1.6 & no data  & 06:39 & 06:49 &  M1.2 & BM\\
\hline
\end{tabular}
\end{table}

\subsection{The 13 June 2012 spike event}

Among these spike events the most interesting examples of spikes were observed
during the 13 June 2012 flare. During this flare we registered many groups of
spikes and also their different types. According to the GOES X-ray observation
this M1.2 flare started at 11:29 UT, maximum at 13:17 UT and ended at 14:31 UT.
In H$\alpha$ observation it started at 11:36 UT, maximum at 14:41 and ended
at 16:23 UT; lasting thus about 5 hours. It occurred at the position S16E18 in
the active region NOAA AR 11504 with an importance 1N.

\begin{figure}    
   \includegraphics[width=1.0\textwidth,clip=]{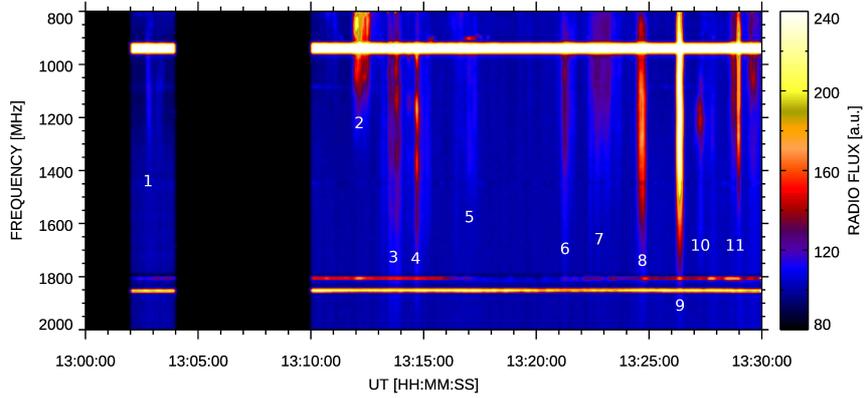}
              \caption{The 800-2000 MHz overview radio spectrum observed at 13:10:00 - 13:30:00 UT
              during the 13 June 2012 flare by the Ond\v{r}ejov radiospectrograph.
              Black parts of radio spectrum at 13:00:00-13:02:00 and 13:04:00-13:10:00 UT
              indicate that we have no data in these intervals because they were not recorded
              owing to no bursts there.}
              \label{fig1}
   \end{figure}

\begin{figure}    
        \center
   \includegraphics[width=0.9\textwidth,clip=]{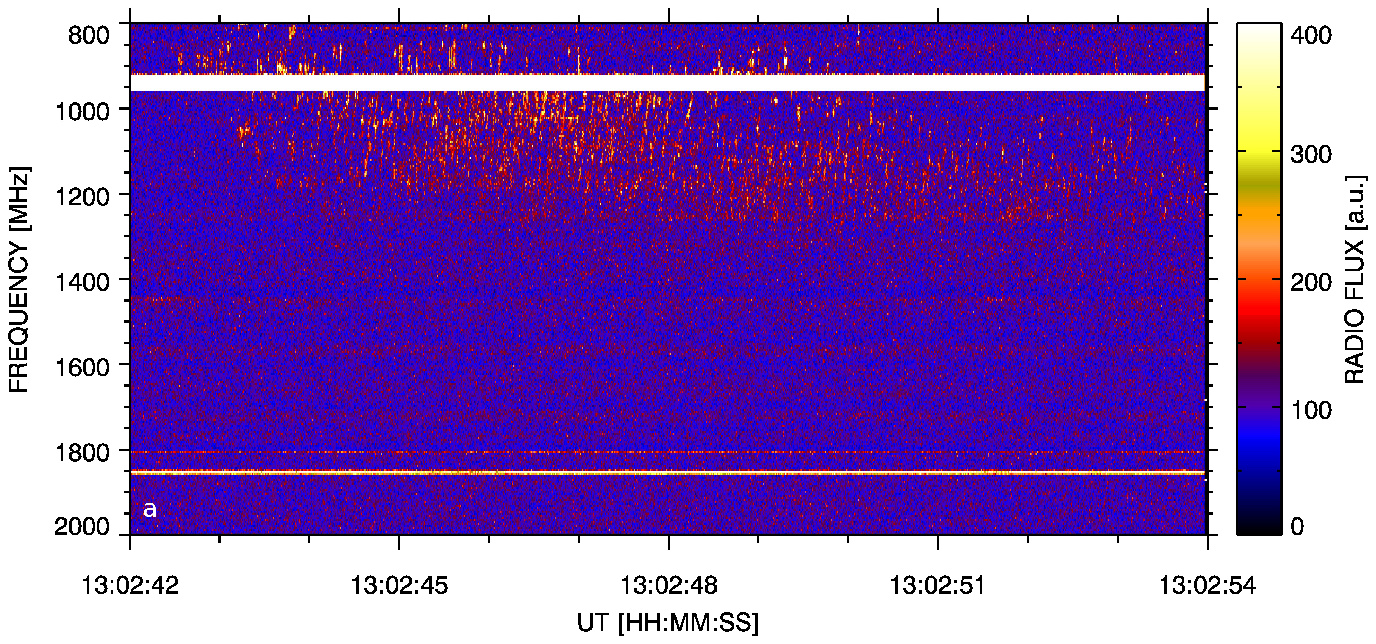}
   \includegraphics[width=0.9\textwidth,clip=]{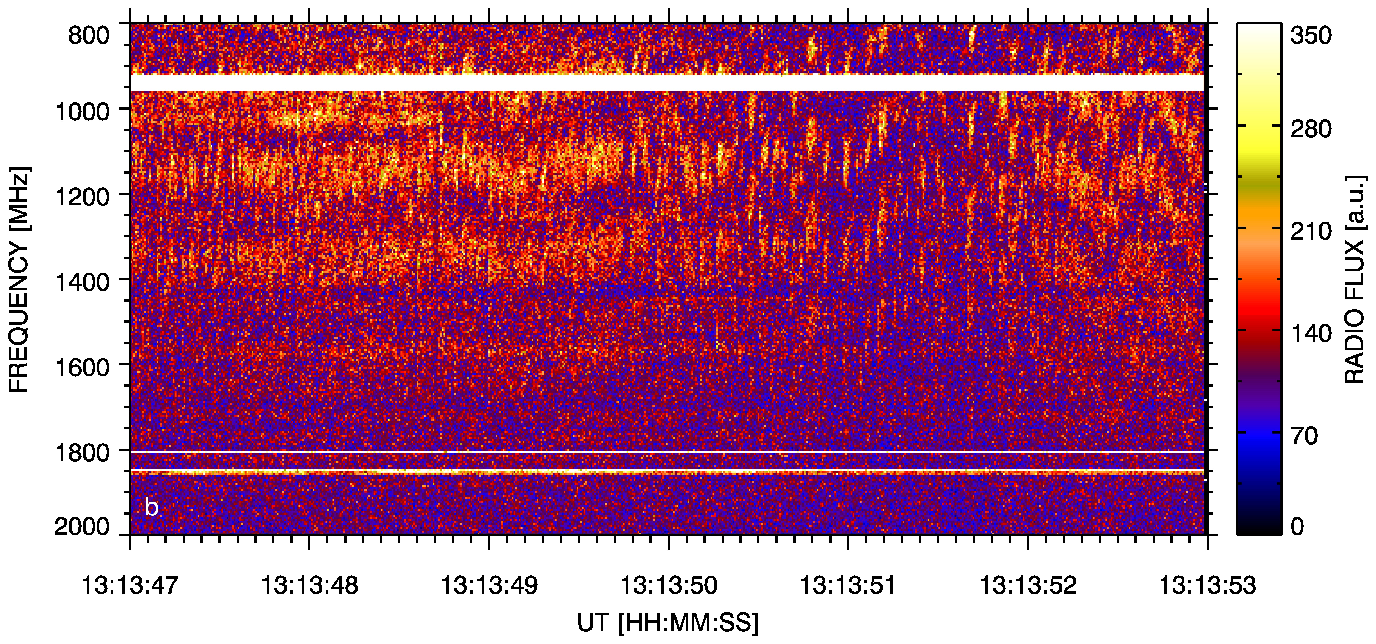}
   \includegraphics[width=0.9\textwidth,clip=]{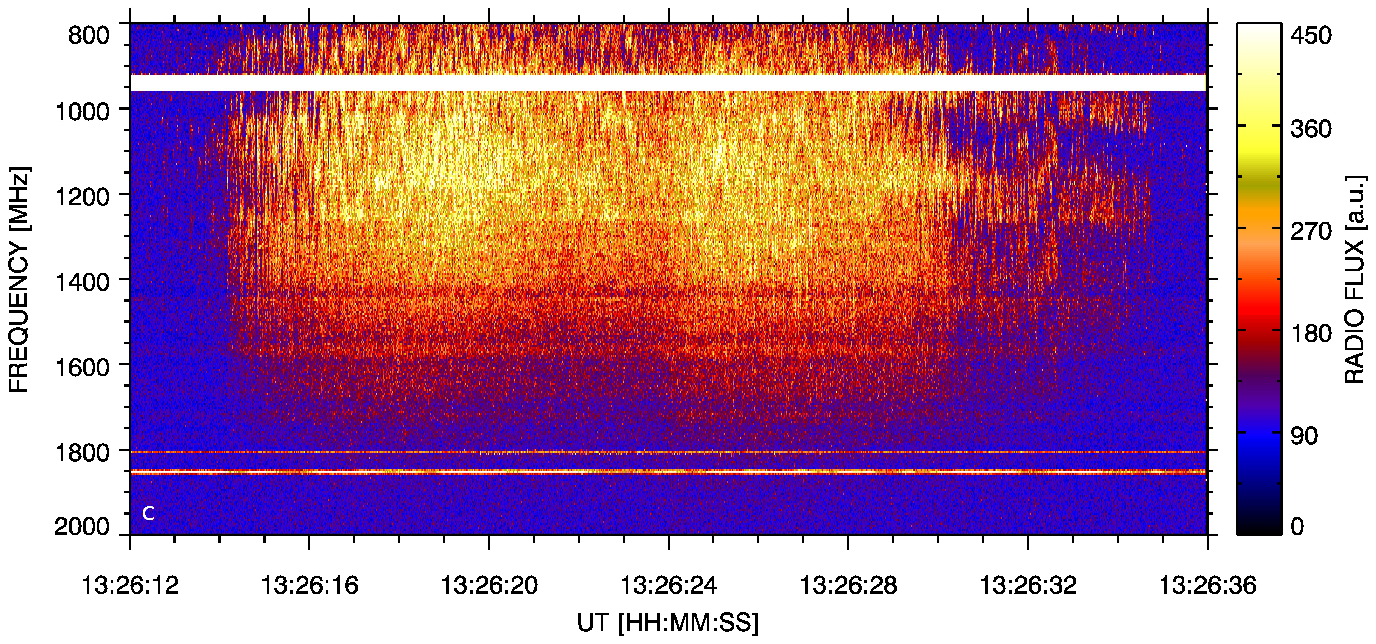}

              \caption{Examples of spikes observed during the 13 June 2012 flare: a) The radio spectrum observed at 13:02:42 - 13:02:54 UT
              showing spikes classified as SB (group 1, see Table~\ref{Table2} and Figure~\ref{fig1}).
              b) The radio spectrum observed at 13:13:47 - 13:13:53 UT
              showing spikes classified as SZ (part of group 3).
              c) The radio spectrum observed at 13:26:12 - 13:26:36 UT
              showing spikes classified as SBN (group 9).}
              \label{fig2}
   \end{figure}

In this flare in the time interval 13:00-13:30 UT we observed eleven groups of
spikes, see Figure~\ref{fig1}. This time interval covers only a part of this
long flare around its GOES X-ray maximum.  Detailed radio spectra of three
groups of spikes (group 1, 3 and 9 in Table~\ref{Table2}) are shown in
Figure~\ref{fig2}. As can be seen here, spikes at these groups differ.
Therefore, we classified them according to their appearance in the radio
spectrum as SB (spikes distributed in broad band or bands)
(Figure~\ref{fig2}a), SZ (spikes distributed in zebra-like bands)
(Figure~\ref{fig2}b), and SBN (spikes distributed in broad and narrow bands)
(Figure~\ref{fig2}c). Owing to resemblance of SZ and zebras, in the following
we will make a comparison of SZ spikes with a typical zebra observed in the
same frequency range (Figure~\ref{fig6}). We will also compare SZ spikes with
SBN spikes.

Using this classification we classified all groups of spikes in the 13 June
2012 flare, see Table~\ref{Table2}. In this table also the basic parameters of
these spike groups are summarized: type, maximal number of the frequency bands
of spikes (MNFB) and characteristic ratios of neighboring bands of spikes (BR).
These ratios were computed from the frequencies at flux maxima of spike bands.
On lower frequencies in the 200-400 MHz range at 13:08-13:24 UT these spike
groups were associated with Type II burst (Callisto-BLEN7M observation).

\begin{table}
\caption{Basic parameters of groups of spikes observed during the 13 June 2012
flare: times, frequency range, types, maximal number of the frequency bands of
spikes (MNFB) and characteristic ratios of neighboring bands of spikes (BR). SB
denotes spikes in broadbands, SZ in zebra-like bands, and SBN
in broad and narrow bands.} \label{Table2}
\begin{tabular}{cccccc}
  \hline
No   & Time & Range & Type & MNFB & BR \\
     &  (UT)& (GHz) &      &      &    \\
  \hline
1 & 13:02:42-13:02:58 & 0.8-1.3 & SB & 1 & - \\
2 & 13:11:48-13:12:40 & 0.8-1.2 & SZ & 2 & 1.21 \\
3 & 13:13:06-13:14:06 & 0.8-1.8 & SZ & 4 & 1.16, 1.19, 1.21 \\
4 & 13:14:38-13:14:50 & 0.8-1.6 & SZ & 4 & 1.15, 1.18, 1.23 \\
5 & 13:16:55-13:17:25 & 0.8-1.5 & SB & 1 & - \\
6 & 13:21:08-13:21:26 & 0.8-1.6 & SB & 1 & - \\
7 & 13:22:21-13:23:18 & 0.8-1.8 & SB & 1 & - \\
8 & 13:24:25-13:24:57 & 0.8-1.8 & SBN & 3 & 1.25, 1.26 \\
9 & 13:26:12-13:26:36 & 0.8-1.8 & SBN & 3 & 1.22, 1.24 \\
10 & 13:26:59-13:27:30 & 1.0-1.6 & SZ & 2 & 1.08 \\
11 & 13:28:55-13:29:05 & 0.8-1.6 & SB & 1 & - \\
\hline
\end{tabular}
\end{table}

\begin{figure}    
\center
\includegraphics[width=0.8\textwidth,clip=]{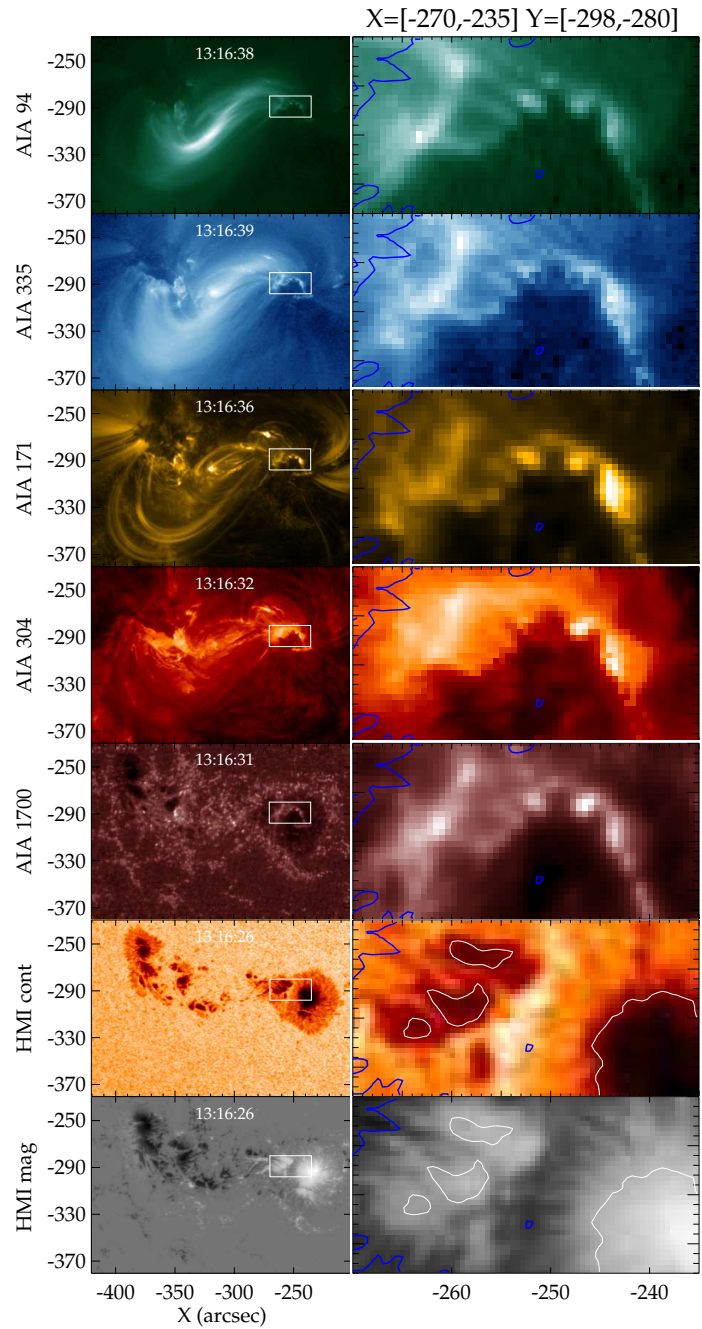}
              \caption{The AIA/SDO and HMI/SDO images of the flare overall region (left) and zoom region (right),
delineated by the white rectangle in the overall images, for five AIA channels
(94, 335, 171, 304, 1700\,\AA) and the HMI continuum and magnetogram at the time close to the
flare maximum (13:16:36 UT).
White isolines in the HMI continuum and magnetogram zoom region images stem for
contours of the leading sunspot umbra and three nearby pores. Blue isolines
denote the neutral division lines between positive and negative magnetic polarities.}
              \label{fig3}
   \end{figure}

\begin{figure}    
\includegraphics[width=1.0\textwidth,clip=]{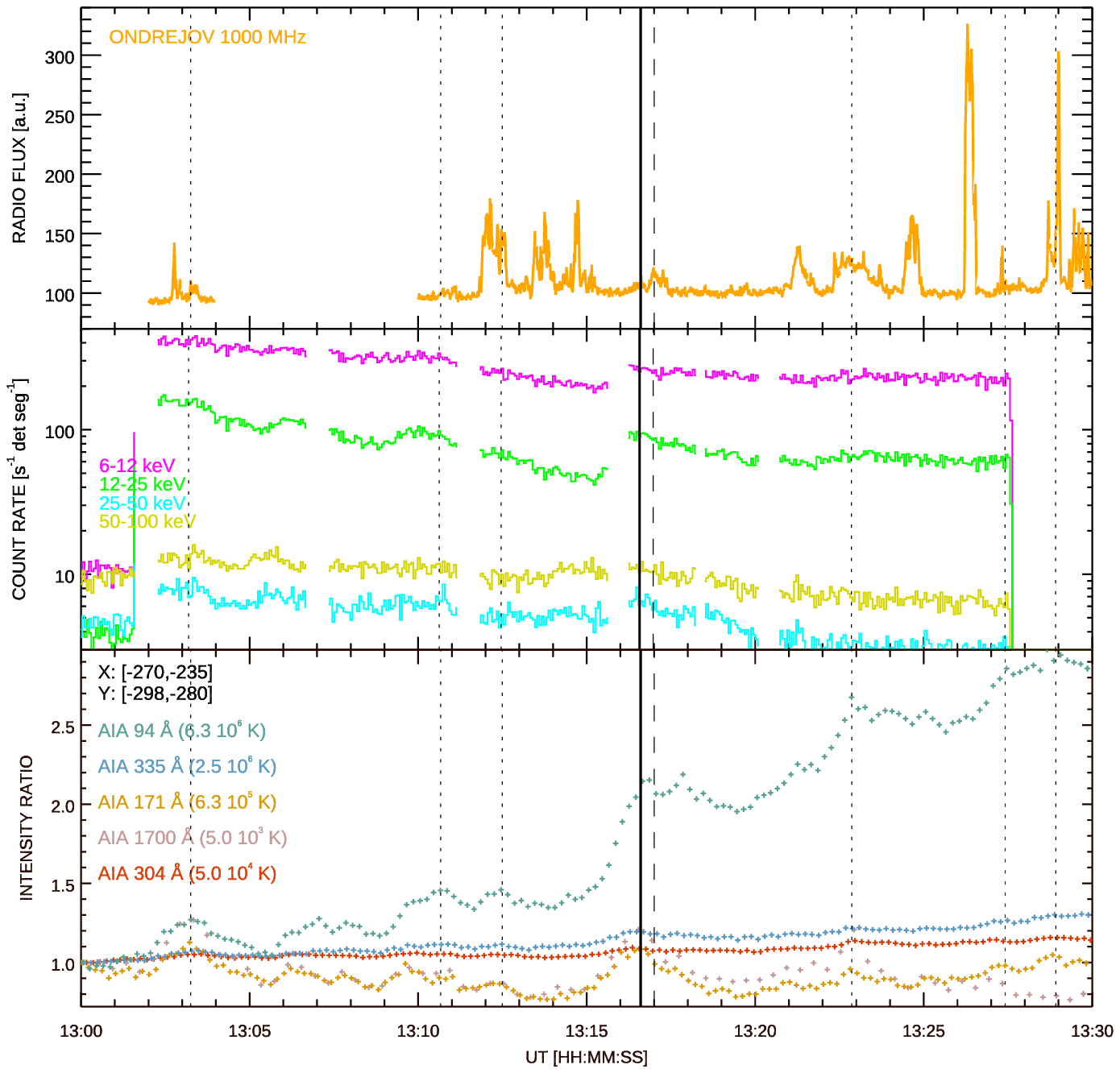}
              \caption{Time profiles in the interval of the radio spectrum (Figure~\ref{fig1}, at 13:00 - 13:30 UT) of the flux at 1000 MHz (upper panel) and RHESSI counts (middle panel) together
 with AIA light curves (bottom panel, expressed as the ratio to the AIA intensity at times around 13 UT) taken from the small white rectangle in Figure~\ref{fig3}.
 The vertical full, dashed and dotted lines in plots of
 time profiles indicate the time of AIA images in Figure~\ref{fig3}, the GOES flare maximum,
 and times of some peaks in the 94 \AA ~time profile, respectively.}
              \label{fig4}
   \end{figure}

AIA/SDO UV \citep{2012SoPh..275...17L}, HMI/SDO
\citep{2012SoPh..275..207S,2012SoPh..275..229S} and RHESSI X-ray
\citep{2002SoPh..210....3L} observations associated with these groups of spikes
are shown in Figure~\ref{fig3} and \ref{fig4}. As seen in Figure~\ref{fig4}
(middle panel), in the time interval of spikes groups, the RHESSI X-rays exhibit
a slow decrease in time with some enhancements; see e.g. that just before the
GOES flare maximum. They indicate a presence of superthermal electrons with the
energies up to 100 keV at these times. Owing to RHESSI technical problems, X-ray
source positions could not be constructed. Therefore, using AIA/SDO
observations we tried to localize a part of the flare where temporal intensity
variations correspond to variations at the 1000 MHz time profile
(Figure~\ref{fig4}). Searching the time profiles in AIA channels (1700, 304,
171, 335 and 94 \AA ~and continuum) in small regions (starting from those with
20 $\times$ 20 arcsec) in the whole flare site we found the most interesting
correspondence with the 1000 MHz radio profile in the white rectangle with the
coordinates X=-270-- -235 and Y=-298 -- -280 arcsec (Figure~\ref{fig3} and
\ref{fig4}). It is interesting that this position corresponds to one end of the
sigmoidal flare structure where many magnetic field lines of flare loops are
concentrated; rooted in the area of the northern part of the active region
leading sunspot, see HMI continuum and magnetogram in Figure~\ref{fig3}. In
other positions AIA profiles were smoother and without such distinct
variations, or with some variations, but only in one AIA channel. As seen in
Figure~\ref{fig4}, there are peaks in several AIA channels observed
simultaneously that correspond to peaks in the 1000 MHz profile. While peaks in
94 \AA~ indicate a presence of a hot plasma (6.3 $\times$ 10$^{6}$ K), the 1700
\AA~ peaks show a heating of deep atmospheric layers with the temperature 5.0
$\times$ 10$^{3}$ K, probably owing to precipitating superthermal particles.
The peaks in radio and AIA are of varying sizes, not all peaks on the 1000 MHz
radio profile have corresponding peaks in AIA channels. It indicates a complex
relation. In fact, a simple relation cannot be expected because it is commonly
assumed that spikes are generated by superthermal electrons with the loss-cone
distribution, but peaks in the AIA channels express plasma density enhancements
at the AIA channel characteristic temperature.

\section{Analysis of spikes, comparison with typical zebra in the same frequency range and modelling of SZ frequencies}

\begin{figure}    
   \centerline{\includegraphics[width=1.0\textwidth,clip=]{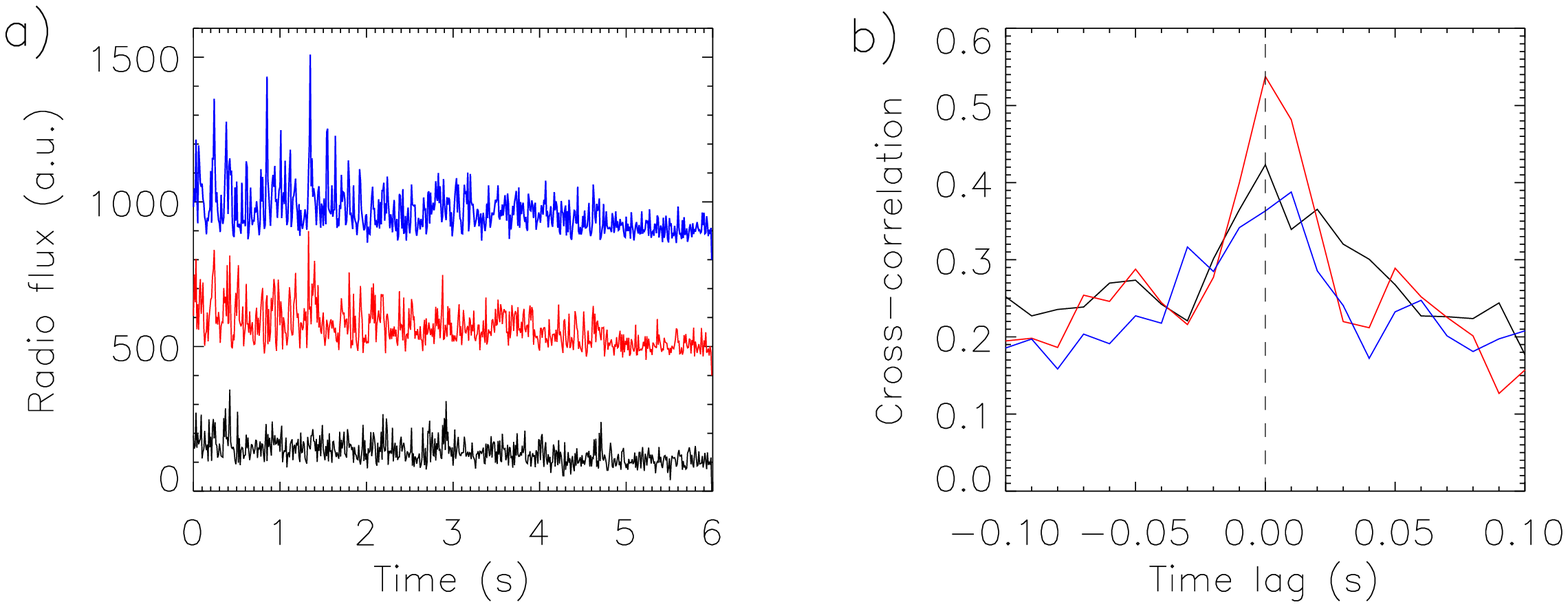}
              }
              \caption{Case with SBN: a) Radio flux vs. time at 1490 MHz (black line), at 1220 MHz
  (red line) + 400 a.u., and at 990 MHz (blue line) + 800 a.u.,
  starting at 13:26:30 UT in the 13 June 2012 event and lasting 6 s. b)
  Cross-correlations of the radio flux profiles at 1490 and 1220 MHz (black line),
  1220 and 990 MHz (red line), and 1490 and 990 MHz (blue line).}
              \label{fig5}
   \end{figure}

\begin{figure}    
             \center
             \includegraphics[width=1.0\textwidth,clip=]{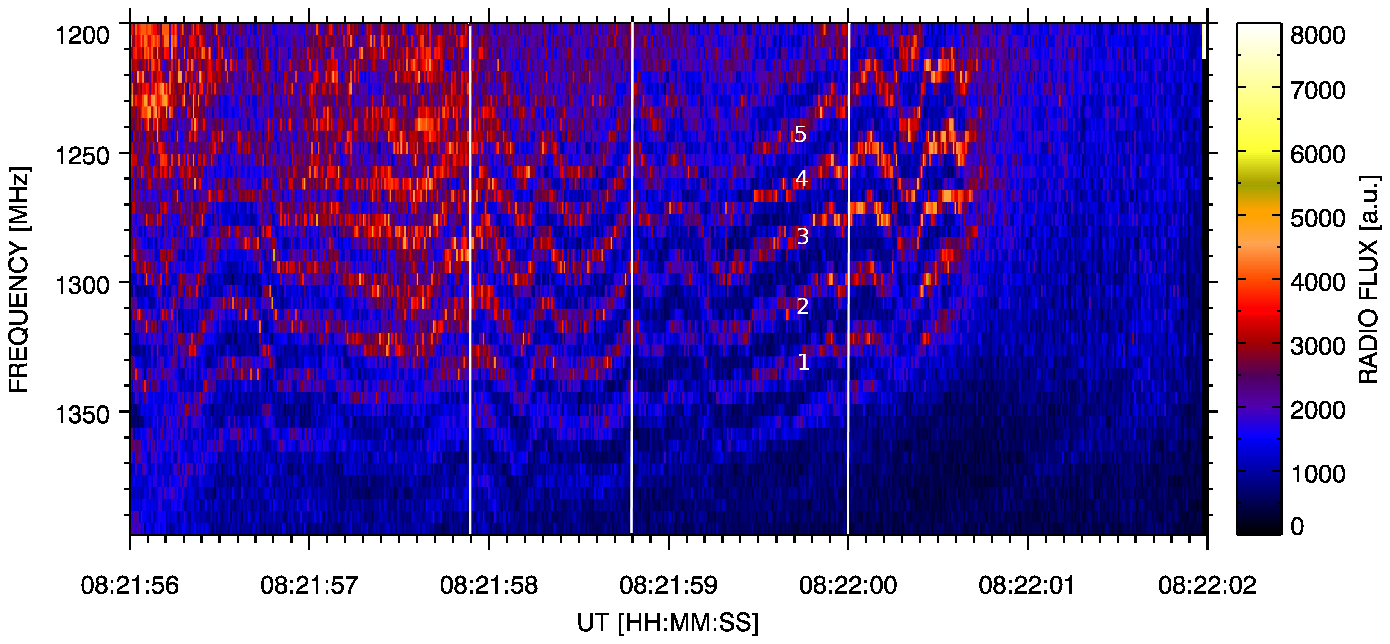}
      \includegraphics[width=0.8\textwidth,clip=]{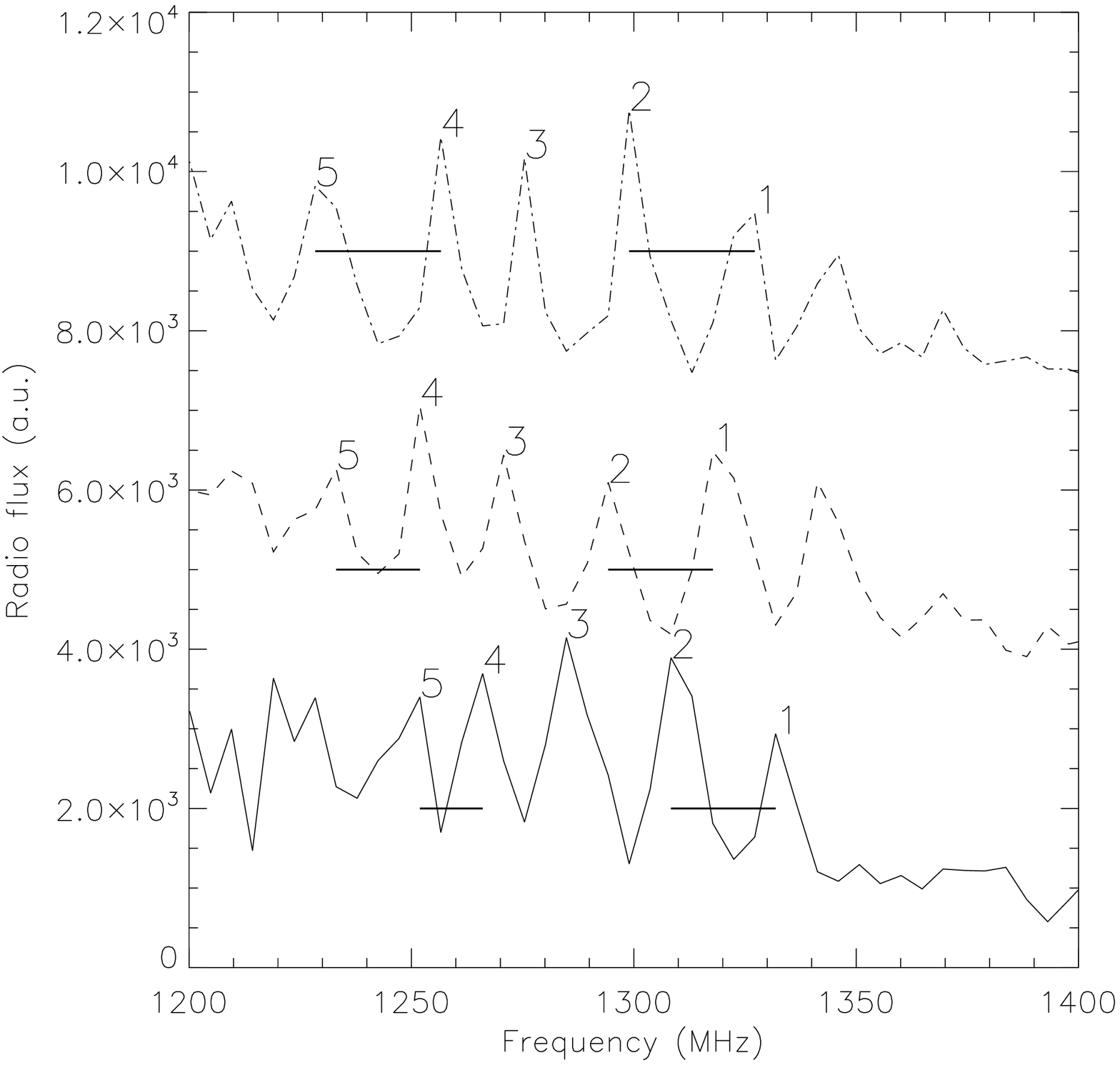}
              \caption{Upper: The 1200-1400 MHz radio spectrum showing the zebra observed during the 1 August 2010 flare. The white vertical lines and numbers 1-5
              show the times of the frequency profiles and selected zebra stripes shown in bottom panel of the Figure.
              Bottom: Radio flux vs. frequency profiles: the full line shows the radio flux at 08:21:57.9 UT, dashed line shows the
              radio flux + 3500 a.u. at 08:21:58.8 UT, and dashed-dotted line shows the radio flux + 7000 a.u. at 08:22:00.0 UT. Numbers 1-5 designate
              zebra stripes; see the upper panel. Short horizontal lines show the frequency intervals between zebra stripes 1 and 2 and zebra stripes 4 and 5 at
              three instants 08:21:57.9, 08:21:58.8 and 08:22:00.0 UT.}
              \label{fig6}
   \end{figure}

Radio observations show three types of spike distributions: a) in one cloud of
spikes (SB), b) in several zebra-like bands (SZ), and c) clouds of spikes with
very narrow bands of spikes (SBN). The presented SBN type of spikes is similar
to the event from 7 November 2013 that was interpreted as emission in
Bernstein modes \citep{2021ApJ...910..108K}. To confirm this similarity we
computed cross-correlation at three spike-band frequencies from the ending part
of SBN: 990, 1220, and 1490 MHz (Figure~\ref{fig5}), where the spike bands are
the narrowest. We found that the time lag is $\leq$ 0.01 s, in agreement with
the 7 November 2013 case. The maximal cross-correlation in Figure~\ref{fig5} is
even higher than that in Figure 3 in the paper by \cite{2021ApJ...910..108K}.
Thus, we think that the present SBN spikes are of the same origin as the spikes
observed at 7 November 2013, i.e., the emission is generated in Bernstein
modes. Furthermore, we compare SB and broad part of SBN spikes. Owing to the
similar parameters of individual spikes we think that the SB type is as one
broad part of SBN type.

   \subsection{Comparison of zebra-like spikes (SZ) with a zebra in the same frequency range}

\begin{figure}    
   \centerline{\includegraphics[width=1.0\textwidth,clip=]{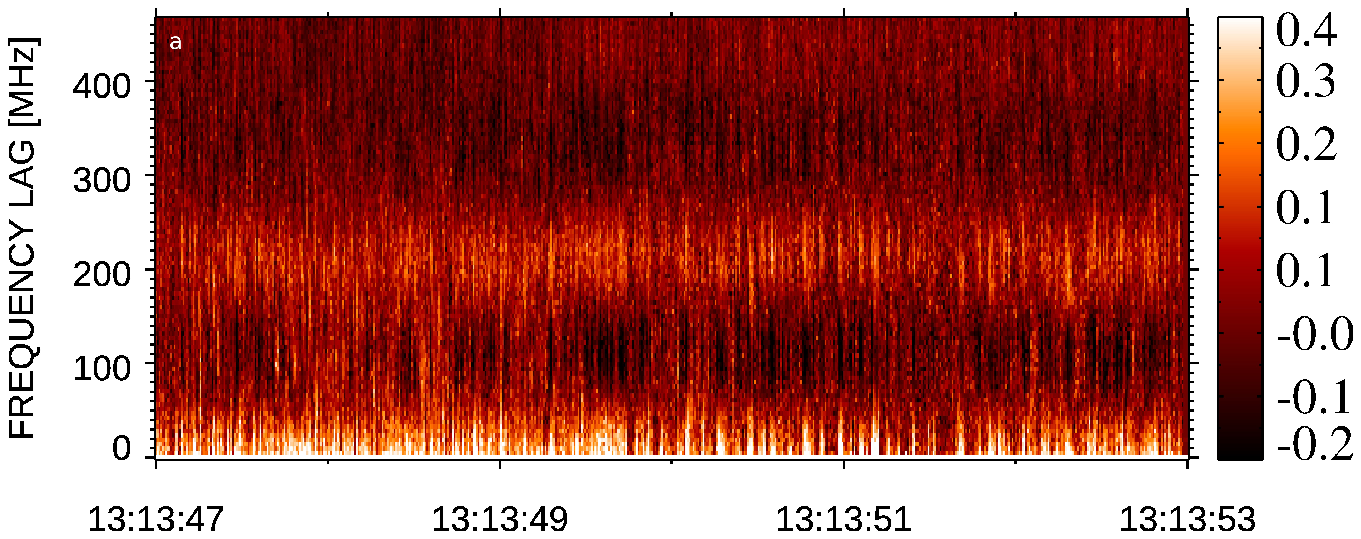}}
   \centerline{\includegraphics[width=1.0\textwidth,clip=]{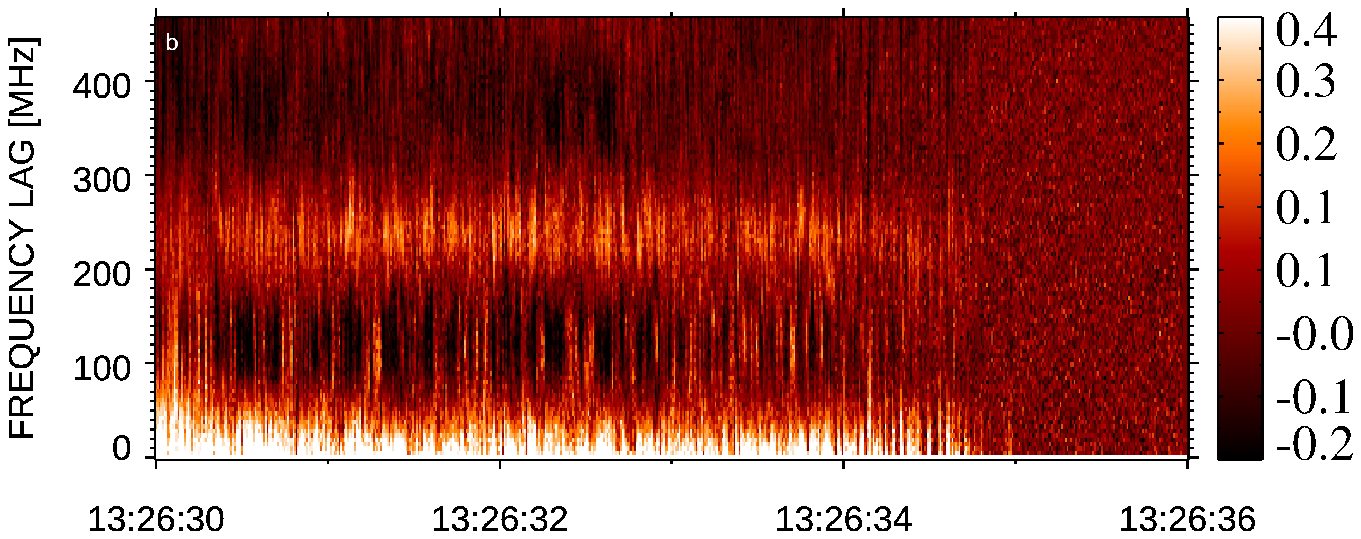}}
   \centerline{\includegraphics[width=1.0\textwidth,clip=]{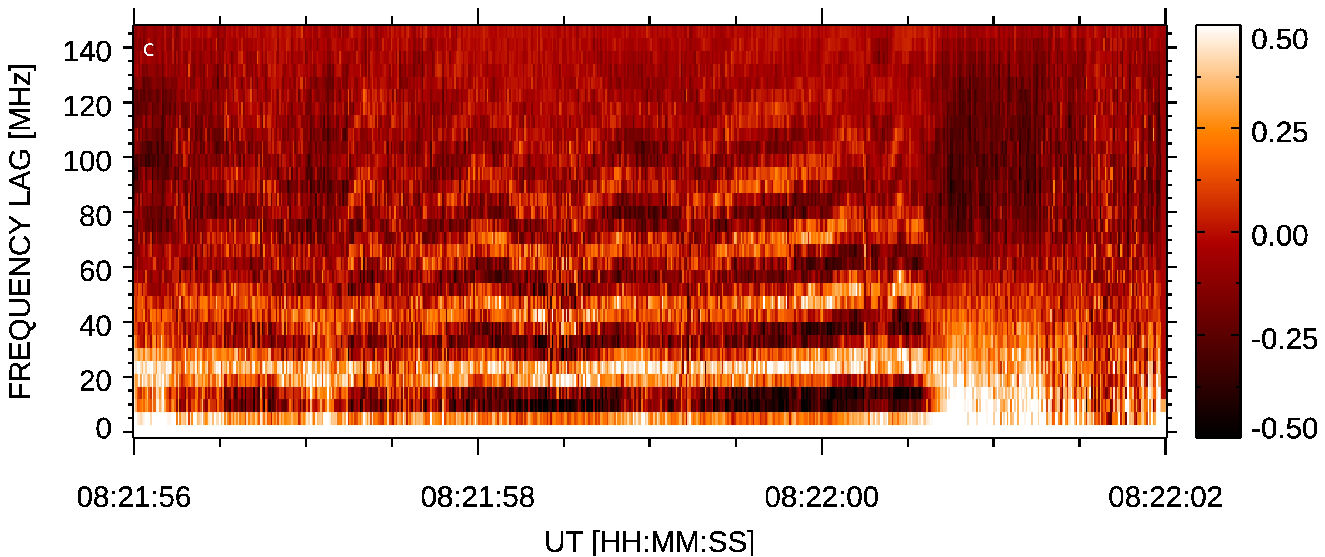}}
              \caption{Auto-correlation of the radio fluxes vs. radio spectrum frequency. a) SZ case in the time interval 13:13:47-13:13:53 UT.
              b) SBN case in the time interval 13:26:30-13:26:36 UT.
              c) Zebra case in the 1200-1400 MHz range observed at 08:21:56-08:22:02 UT on 1 August 2010.}
              \label{fig7}
   \end{figure}

The most interesting type of spikes is the SZ type because it resembles zebras.
Therefore, let us compare it with a typical zebra in the same frequency
range, e.g., with the zebra observed at 1 August 2010 (Figure~\ref{fig6} upper
panel). First, we use the auto-correlation method. We computed the time evolution
of the auto-correlations at different frequencies. We made these computations for SZ
shown in Figure~~\ref{fig2}b and also for the ending part of SBN in
Figure~\ref{fig2}c and the zebra shown in Figure~\ref{fig6} upper panel. The
results are in Figure~\ref{fig7}. The auto-correlations of type SZ and SBN are
similar. Their frequency lag is about 220 and 250 MHz, respectively, and this
lag is practically constant in time. When we divide the mean SZ band
frequencies ($\sim$960, $\sim$1120, $\sim$1330 and $\sim$1550 MHz) by the lag
frequency ($\sim$220 MHz) the ratios $\sim$4.4, $\sim$5.1, $\sim$6.0 and
$\sim$7.0 are obtained. Similarly, dividing the mean SBN band frequencies at
the ending part of SBN ($\sim$990, $\sim$1220 and $\sim$1490 MHz) by the lag
frequency ($\sim$250 MHz) the ratios $\sim$4.0, $\sim$4.9 and $\sim$6.0 are
obtained. On the other hand, the auto-correlations of the zebra differ
significantly. The frequency lag at first harmonic is about 24 MHz. When the mean
zebra frequency ($\sim$ 1250 MHz) is divided by the lag frequency ($\sim$ 24
MHz) then this ratio is about 52.

Moreover, the frequency lag in the zebra case is varying in time as seen in
higher harmonics of time lags. To understand better what it means, we
calculated the frequency intervals between neighboring zebra stripes at three
time instants (Figure~\ref{fig6} bottom panel). As shown here, for the zebra
stripes numbered 1 and 2 the frequency interval between them (see the short
horizontal lines) changes from 23.5 MHz to 28.2 MHz, and for
the zebra stripes numbered 4 and 5 the frequency interval between them changes
from 14.1 MHz to 18.8 MHz and to 28.2 MHz. (Note that the frequency resolution
of the radio spectrum is 4.7 MHz.) It means that the frequency interval between
the neighboring zebra stripes changes and this frequency interval can be
different from those between other zebra stripes. This irregular variation of
the separation frequency indicates that zebra-stripe sources cannon be
generated in Bernstein modes in one source region. In the double-plasma
resonance models of zebras this variation is interpreted by waves propagating
along the loop where the zebra-stripe sources are located
\citep{2021A&A...646A.179K}. We also found that the bandwidth of SZ bands is
much broader than that of zebra stripes. While in the SZ case it is about 200
MHz in the zebra case it is about 10 MHz.

\subsection{Modelling of SZ frequencies by Bernstein modes}

\begin{figure}    
    \centerline{\includegraphics[width=1.0\textwidth,clip=]{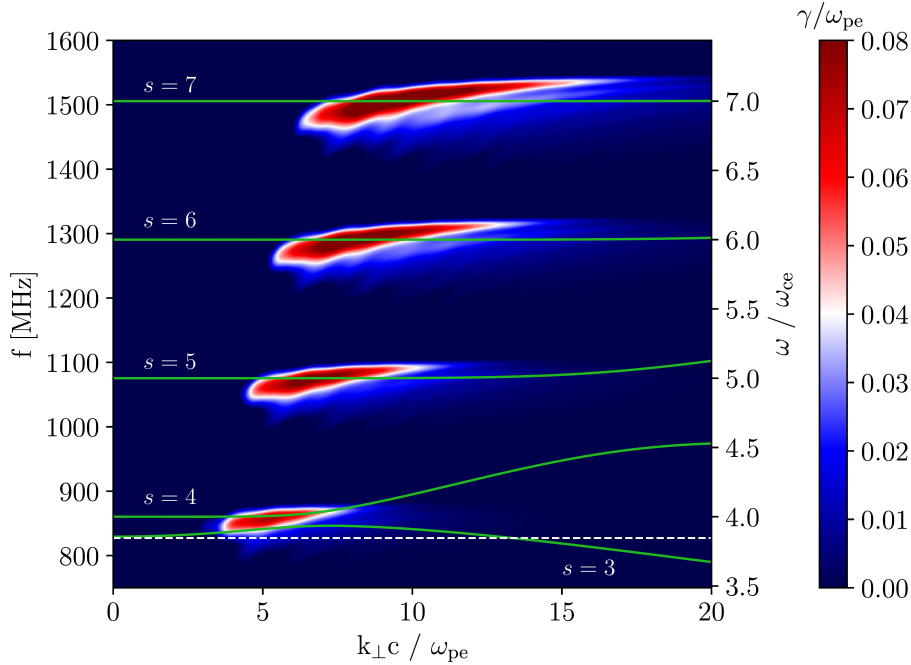}}
          \caption{Growth rates of the Bernstein modes as a function of the frequency
  and perpendicular k-wave vector for parameters $\omega_\mathrm{pe}/\omega_\mathrm{ce}=3.85,
  v_\mathrm{tb}/c=0.01834, v_\mathrm{t}/c=0.25, n_\mathrm{e}/n_\mathrm{h} = 15$.
  The values of f in MHz correspond to the radio emission in Bernstein modes, i.e. $\omega_\mathrm{pe}= 2 \pi f_\mathrm{pe}$, $f_\mathrm{pe}
= 827$~MHz.
   Green lines:
  Dispersion branches computed from the Equation~\ref{eq1}, $s$ is the gyro-harmonic number of each branch.
  White dashed horizontal line: Plasma frequency.}
              \label{fig8}
   \end{figure}

To confirm a similarity between SZ and SBN of the 13 June 2012 flare given by
auto-correlations (Figure~\ref{fig7}) and SBN observed in 7 November 2013 with
the Bernstein modes, we fitted the mean SZ band frequencies by the same way as
in the SBN 7 November 2013 case \citep{2021ApJ...910..108K}, see also
\cite{2019ApJ...881...21B}. Firstly, we searched for the time where the
bandwidth of SZ spike bands is the narrowest. We found it for 13:13:50.8 UT
where we determined the mean spike band frequencies as 860, 1070, 1290 and 1510
MHz (for $s$ = 4 - 7). Then using Equations~\ref{eq1} and \ref{eq2} we fitted
these band frequencies by calculating the dispersion curves of the Bernstein
modes and corresponding growth rates and assuming that Bernstein modes
correspond to observed radio frequencies. The best fit is shown in
Figure~\ref{fig8}. This fit was obtained with the following parameters:
$n_\mathrm{e}/n_\mathrm{h} = 15, v_\mathrm{t}/c=0.25, v_\mathrm{tb}/c =
0.01834~(T_b = 2 MK), \omega_\mathrm{pe}/\omega_\mathrm{ce}=3.85$, and
$f_\mathrm{pe} = 827$~MHz, where $f_\mathrm{pe} = \omega_\mathrm{pe}/(2 \pi)$,
$n_\mathrm{e}/n_\mathrm{h}$ is the ratio between the electron plasma densities
of the background and hot component plasma, $v_\mathrm{t}$ and $v_\mathrm{tb}$
are the characteristic velocities of hot and Maxwellian background plasmas,
$T_b$ is the temperature of the background plasma, and c is the light speed.

Now, considering the frequency separation between spike bands as follow from
the above analysis (220 MHz) and the plasma frequency as $f_\mathrm{pe} =
827$~MHz, the magnetic field and plasma density in the spike source can be
estimated as about 79 G and 8.4 $\times$ 10$^{9}$ cm$^{-3}$, respectively.

A question also arises what is the expected bandwidth of spikes in the model
of Bernstein modes. But, it is a very complex problem. Therefore, let us
estimate only the Bernstein mode bandwidth for several gyro-harmonic numbers
$s$, assuming that this bandwidth is formed only by density variations. We take
these variations as $\Delta n / n \approx 1 / \sqrt{n} \approx 1.1\times
10^{-5}$ (statistical noise). Then, the frequency variation at the present
plasma frequency (827 MHz) is about 2.7 MHz. We estimated the linearized
bandwidths in region of each branch, where its growth rate is maximal. We found
the bandwidth of Bernstein modes with $s = 4, \ldots, 7$ as 0.9, 3$\times$
10$^{-3}$, 5$\times$ 10$^{-5}$ and 5$\times$ 10$^{-5}$ MHz, respectively. The
bandwidths narrow as the permittivity derivation nonlinearly increases with
increasing of the gyro-harmonic number. For all $s$, the Bernstein mode
bandwidth is smaller than the minimal spike bandwidth found by
\cite{2014ApJ...789..152N} ($\sim$ 1 MHz). However, the problem of the spike
bandwidth is much more complex. It is due to not only effects of density and
magnetic field variations, but also due to the conversion of Bernstein modes
into electromagnetic waves. Moreover, density and magnetic field variations
are probably interconnected.

\section{Discussion and conclusions}

We confirmed that the narrowband dm-spikes are mostly observed during the
impulsive flare phases, see Table~\ref{Table1}. As follows from
Table~\ref{Table2} and analysis of the 13 June 2012 spike event, dm-spikes are
observed in bands with non-integer ratio in the range 1.08--1.26; in
agreement with the result of \cite{1994A&A...285.1038K}. We note that this
result is not frequently considered in theoretical models of dm-spikes.

In the analysis of the 13 June 2012 spike event we tried to search for some
relation between the 1000 MHz time profile (with groups of spikes) and
variations of intensities in selected flare locations using AIA/SDO observations.
The most interesting relation was found for AIA intensities taken from one end
of the sigmoidal flare structure where many magnetic field lines of flare loops
are concentrated (Figure~\ref{fig3} and \ref{fig4}).  The peaks in radio and
AIA (Figure~\ref{fig4}) are of varying amplitudes, not all peaks on the 1000 MHz
radio profile have corresponding peaks in AIA channels. It indicates that their
relation is not simple. In reality, a simple relation cannot be expected
because spikes are generated by superthermal electrons with a loss-cone
distribution while peaks in the AIA channels indicate plasma density
enhancements at the AIA channel characteristic temperature.

During one flare from 13 June 2012 we observed spikes with three different
types: SB, SZ and SBN spikes. Analyzing SBN spikes in their narrowband part by
the cross-correlation method we found that the SBN spikes are similar to those
presented in the paper by \cite{2021ApJ...910..108K}. Thus, we interpret SBN
similarly, i.e., as generated in Bernstein modes. We also found a similarity in
autocorrelations of SZ and SBN ending part which speaks in favor of the same
generation mechanism for these spikes.

We made a detailed comparison of the 13 June 2012 SZ (zebra-like) spikes with
a typical zebra in the same frequency range that was observed at 1 August
2010. We found the following differences. In the SZ case: The separation
frequency $\Delta f$ between neighboring spike bands is about 220 MHz. The
autocorrelation in SZ indicates low variability in time. The ratios between
four spike-bands and separation frequencies was found as 4.4, 5.1, 6.0 and 7.0.
Similarly, in the SBN case these ratios for three spike bands are 4.0, 4.9 and
6.0.  On the other hand, in the zebra case: The separation frequency $\Delta f$
between neighboring zebra stripes is about 24 MHz.  The variability of the
autocorrelation in time in the zebra case is higher than in SZ case.
Furthermore, the ratio between the mean zebra-stripe frequency and separation
frequency is about 1250 MHz/24 MHz $\sim$ 52.

Moreover, we found that the separation frequency between neighboring zebra
stripes changes by different way for different pairs of the neighboring zebra
stripes (Figure~\ref{fig5}). This irregular variation of the separation
frequency excludes that the analyzed zebra is generated in one source as is in
the model of the Bernstein modes.

The bandwidth of of SZ spike bands (consisting many narrowband spikes) is much
broader than that of zebra stripes. For this reason it was impossible to make
a similar analysis of time evolution of band (or spike) separation
frequencies as in the zebra case shown in Figure~\ref{fig6} bottom panel.

Nevertheless, at time of the narrowest SZ bands we successfully fitted SZ
frequencies by the model of the Bernstein modes. Using the parameters in this
model we estimated the mean magnetic field strength and plasma density in the
SZ source as about 79 G and 8.4 $\times$ 10$^{9}$ cm$^{-3}$, respectively. We
propose that broad bands of SZ spikes correspond to a region with some interval
of the magnetic field and density. In accordance with our previous ideas
\citep{1996SoPh..168..375K,2001A&A...379.1045B} we think that this region is in
the magnetic reconnection outflow, where the plasma is in turbulent state.

Considering all these facts, we conclude that SZ and SBN spikes observed on 13
June 2012 were generated according to the model of the Bernstein modes. We
believe that the parameters, e.g. the ratio between the band (stripe) and
separation frequencies, found in the analysis of SZ spikes and the 1 August
2010 zebra, can help in interpretations of other observed SZ spikes and zebras.

In Section 3.2 we tried to estimate the spike bandwidth in the model of
Bernstein modes. We considered only the Bernstein mode bandwidth assuming only
an effect of the density variations. But this problem is much more complex. It
is due to not only effects of density and magnetic field variations, but also
due to the conversion of Bernstein modes into electromagnetic waves.
Moreover, density and magnetic field variations are probably interconnected.
Therefore, the question about the spike bandwidth remains open and further
theoretical analysis is necessary.

Similarly as in other solar radio bursts, the radio-wave scattering can play a
role also in spikes. The scattering is proportional to (f$_{\rm pe}$/f)$^2$
\citep{1993ASSL..184.....B}, where f$_{\rm pe}$ is the plasma frequency and f
is the frequency of the radio emission. The scattering depends on the size of
the scattering region, sizes and density irregularities. The scattering
enhances sizes of the radio sources and causes temporal smoothing of time
variations of the burst intensity. Moreover, there is wave ducting
of the radio emission \citep{1979SoPh...63..389D} that requires "fibrous"
structures of the plasma density \citep{2020ApJ...898...94K}. To provide the
scattering rate comparable with that of the observations, magnetic tubes would
need to have a density contrast of $\Delta$n/n $\gg$ 1 (e.g.
\cite{1983PASA....5..208R} request a 25-fold increase of the plasma density
over dense fibres in the ducting model). Existence of such structures in the
solar corona is not supported by EUV observations
\cite[e.g.][]{2020ApJ...890...75M}. So the "ducting" model is quantitatively
inconsistent with the observations and anisotropic scattering is required.

In the present study without spike source imaging, the scattering can only
smooth and prolong duration of spikes. But, the parameters of the scattering
region are not known. We found that the plasma frequency for the narrowest
spike bands at 13:13:50.8 UT is  827 MHz and Bernstein mode frequencies  are
860, 1070, 1290, and 1510 MHz. Because the lowest Bernstein frequency 860 MHz
is 33 MHz  above the plasma frequency that is why  the scattering of spike
emission in the model of Bernstein modes is less important than the
scattering in the case of the radio emission at the fundamental frequency (f
$\sim$ f$_{\rm pe}$) as e.g., in the type III bursts. For Bernstein modes with
increasing $s$ the effect of scattering decreases.

\begin{acks}
M. K. acknowledges support from the project RVO-67985815 and GA \v{C}R grants
20-09922J, 20-07908S, 21-16508J and 22-34841S. J.R. support by the Science
Grant Agency project VEGA 2/0048/20 (Slovakia), J.B. support by the German
Science Foundation (DFG) project BU 777-17-1, and J. K. support by GA \v{C}R
grant 19-09489S. Help of the Bilateral Mobility Project SAV-18-01 of the SAS
and CAS is acknowledged as well. We also acknowledge the use of the Fermi Solar
Flare Observations facility funded by the Fermi GI program. Data supplied
courtesy of the SDO/HMI and SDO/AIA consortia.
\end{acks}

\bibliographystyle{spr-mp-sola}
\bibliography{spi-bands}

\end{article}

\end{document}